\documentclass[%
 aip,
 amsmath,
 amssymb,
 reprint,%
]{revtex4-1}

\usepackage{xcolor}

\usepackage[utf8]{inputenc}
\usepackage[T1]{fontenc}
\usepackage{mathptmx}
\usepackage{etoolbox}
\usepackage{iftex}
\usepackage{CJKutf8}

\usepackage{graphicx}
\usepackage{dcolumn}
\usepackage{bm}
\usepackage[utf8]{inputenc}
\usepackage{mathptmx}
\usepackage{etoolbox}
\usepackage{mathrsfs}
\makeatletter
\def\@email#1#2{%
 \endgroup
 \patchcmd{\titleblock@produce}
  {\frontmatter@RRAPformat}
  {\frontmatter@RRAPformat{\produce@RRAP{*#1\href{mailto:#2}{#2}}}\frontmatter@RRAPformat}
  {}{}
}%
\makeatother
\begin{document}
\preprint{AIP/123-QED}
\title[Flow Completion Network]{Flow Completion Network: Inferring the Fluid Dynamics from Incomplete Flow Information using Graph Neural Networks}

\author{Xiaodong He\begin{CJK*}{UTF8}{gbsn}
（何校栋）\end{CJK*}}
  \affiliation{Department of R\&D, UnionString(Beijing) Technology Co. Ltd., Beijing, CHINA.}
\author{Yinan Wang\begin{CJK*}{UTF8}{gbsn}
（王亦南）\end{CJK*}}
  \affiliation{Department of Mechanical, Material and Aerospace Engineering, University of Liverpool, Liverpool L69 3BX, UK.}
\author{Juan Li\begin{CJK*}{UTF8}{gbsn}
（李娟）\end{CJK*}}
  \email[Corresponding Author: ]{juan.li@kcl.ac.uk}
  \affiliation{Department of Engineering, King's College London, London WC2R 2LS, UK.}

\date{\today}
\begin{abstract}
This paper introduces a novel neural network - \textcolor{black}{flow completion network (FCN)} - to infer the fluid dynamics, including the flow field and the force acting on the body, from the incomplete data based on Graph Convolution Attention Network. The FCN is composed of several graph convolution layers and spatial attention layers. It is designed to infer the velocity field and the vortex force contribution of the flow field when combined with the vortex force map (VFM) method. Compared with other neural networks adopted in fluid dynamics, the FCN is capable of dealing with both structured data and unstructured data. The performance of the proposed FCN is assessed by the computational fluid dynamics (CFD) data on the flow field around a circular cylinder. The force coefficients predicted by our model are validated against those obtained directly from CFD. Moreover, it is shown that our model effectively utilizes the existing flow field information and the gradient information simultaneously, giving a better performance than the traditional \textcolor{black}{convolution neural network (CNN)-based and deep neural network (DNN)-based} models. \textcolor{black}{Specifically, among all the cases of different Reynolds numbers and different proportions of the training dataset, the results show that the proposed FCN achieves a maximum norm mean square error of 5.86\% in the test dataset, which is much lower than those of the traditional CNN-based and DNN-based models (42.32\% and 15.63\% respectively).} 
\end{abstract}

\maketitle


\section{\label{sec:Introduction}Introduction}
Flow field completion and body force extraction from incomplete flow field information are important in a range of applications. An example is in one of the key approaches in experimental fluid dynamics - particle image velocimetry (PIV) \citep{Lagemann2021,Wen2019}, where non-invasive force measurements have long been a challenging task. Another application is the load prediction and controls in aeroelastic problems, such as the wind farm flow prediction and control from LIDAR measurements \citep{ZHANG2021116641}. The flow field reconstruction from sparse sensors \citep{Fukami2021} also involves the flow completion techniques. 
Solving the Navier-Stokes equations in computational fluid dynamics (CFD) could provide the detailed flow field and the pressure distribution and skin friction on the body surface, thus the unsteady force acting on the body. Direct load measurements in PIV can be significantly contaminated by resonance effect \citep{DeVoria2014,Ram2014}, and it can be advantageous to obtain force information instead by computing them from the measured flow field. Resolving these forces directly from surface pressures and skin friction has been challenging since resolving the entire boundary layer to an adequate resolution near the solid surface is not realistic in most experimental measurements \citep{DeVoria2014}. Instead, volumetric pressure-free methods \citep{wu1981,noca1996,chang1992,Howe1995,Kang2018} achieve success in taking advantage of accurate experimental measurements of flow fields such as PIV, and extracting the force on the body in a non-intrusive way. \citet{LIWU2018} proposed a vortex force map (VFM) method by further exploring \citet{Howe1995}'s force formula for the derivation of the body force from the vorticity field. The VFM method has been well extended to a finite and limited chosen region enclosing the body \citep{Lietal2021} and to three-dimensional flows \citep{li_zhao_graham_2020}. However, these methods still require detailed flow field information at least in a specific domain enclosing the body. Therefore, in this work, we will explore the flow field reconstruction from very limited incomplete measurements and further predict the body force combined with the previously proposed VFM method. 

The recent boom of data-driven approaches and the proliferation of high-quality experimental or CFD flow data have attracted great attention in data-driven inference to simulate, reconstruct, or predict the fluid dynamics properties. Currently, high-fidelity CFD simulation is still resource-intensive and limits their use in industrial applications requiring quick turnarounds. Low-order theoretical descriptions of flow features has seen some success with analytical-numerical coupling methods \citep{Ram2014,LIWU2016} but are very limited in their scope of application. The situation is now changing with the introduction of machine learning, which has been widely used in reconstruct or the surrogate modeling of the flow fields according to the information collected from either experiments or numerical simulations \citep{Maulik2020,CaiEtal2019}. 
\citet{CNN_SuperResolution_Kai_2019} adopted a standard convolution neural network (CNN) and developed an improved hybrid Downsampled Skip-Connection Multi-Scale (DSC/MS) model to reconstruct the high-resolution flow field from grossly under-resolved flow field data. They showed a remarkable ability to reconstruct laminar and turbulent flow fields from low-resolution data. \citet{Morimoto2020} developed a CNN-based method to estimate the velocity field through imperfect experimental (PIV) measurements of snapshots with missing data. \citet{CNN_CFDAcc_Kochkov_2021} use end-to-end CNN based model to improve approximations inside computational fluid dynamics for modeling two-dimensional turbulent flows. Their research exemplifies how scientific computing can leverage machine learning and hardware accelerators to improve simulations without sacrificing accuracy or generalization. \citet{DNN_DeepVIV_Raissi_2019,DNN_HFM_Raissi_2020} developed a classical physical-informed deep neural network by including the N-S equation in the loss functions to infer the velocity, pressure, and hence the lift and drag from limited and scattered time-space data of the velocity field. This method predicted precisely the flow information within the range of the training set. \citet{Miyanawala2017AnED} proposed an efficient model reduction technique based on CNN and the stochastic gradient descent method to predict the unsteady fluid dynamic forces for different geometry at low Reynolds numbers.
\citet{Bhatnagar2019} built a surrogate model for flow field prediction based on CNN, which was shown to predict the velocity and pressure field orders of magnitude faster than the RANS solver. Specific convolution operations, parameter sharing, and gradient sharpening are used to improve the capability of the CNN.

Most of the traditional CNN-based methods are inherently limited to utilizing structured data since these methods need a generation of a feature matrix that could not apply to unstructured data. Flow measurements are however highly unstructured or even scatter distributed. Moreover, the standard CNN is translation invariant and sensitive to the scale of the data. The resolution of the output data depends on the scale of the training data and the resolution of the input data.
In addition , it is difficult to directly use the N-S equation as a loss function in a standard CNN model since CNN-based model structure could not automatically calculate the partial derivative of coordinates through existing deep learning frameworks (such as tensorflow and pytorch). \textcolor{black}{Other works \citep{Ali2021,Ali2020PhysicsInformedPA} on deep learning of CFD on irregular geometries and unstructured grids have overcome the limitations of CNN for complex geometries for steady flow problems.}

To perform inference on unstructured or mesh-free data, \citet{Trask2020} introduced GMLS-Nets, which parameterize the generalized moving least-squares functional regression method. The GMLS-Nets demonstrated successful prediction of body forces on a cylinder dataset based on unstructured point cloud fluid data. \citet{Ogoke_2021} proposed a data-driven graph neural network (GNN) framework, extended from GraphSAGE~\citep{GNN_GraphSAGE_William_2017}, for the drag force prediction of flow field from irregular and unstructured data. In \citet{GNN_GraphSAGE_William_2017}'s work, Top-K pooling step is used to replace the feature aggregation in \citet{GNN_GraphSAGE_William_2017}. Whereas, non of these existing GNN-based models apply the laws of physics (NS equations in the fluid dynamics) to the flow field prediction, which is proved to be vital to the accuracy, efficiency, and generalization capability of the model \citep{Wandel2021}. Thus, physical-informed GNN applied to unstructured data needs further exploration.

In this work, a novel deep learning model - flow completion network (FCN) - updated from the GraphSAGE \citep{GNN_GraphSAGE_William_2017}, is designed to accurately predict the velocity field from an incomplete knowledge of the existing flow field data. Combined with the VFM method, the predicted velocity field is directly used to infer the force contribution of the vortex flow field, avoiding the utilization of an intermediate variable - the pressure field. It is well understood that the over-smoothing~\citep{Graph-OverSmooth-1,Graph-OverSmooth-2} and the lack of gradient information are detrimental to the convergence rate and the accuracy of the GNN models. Thus, in our FCN, 5 neural net work layers are introduced instead of using a deeper GNN in order to supress the over-smoothing phenomenon ~\citep{Graph-OverSmooth-3,Graph-OverSmooth-4,Graph-OverSmooth-5}. The 5 neural net work layers consist of three graph convolution (GC) layers and two spatial gradient attention (SGA) layes, where the SGA layer could also utilize the gradient information between the reference nodes to performs an accurate information transmission and flow field prediction. Unlike the traditional CNN model which has limited application in structured data, this model is free from the constraints of data structure and could deal with both structured and unstructured data. 

To effectively use the gradient information between nodes, gradient attention layers are carefully designed to facilitate the transmission of gradient information between nodes. This procedure greatly simplifies the structure of the model and increases its performance. Moreover, the N-S equation is integrated into the GNN model training as loss functions, to make sure the obtained model conforms to the physical laws. 

The experimental results show that the proposed FCN model could accurately predict the flow features such as the velocity in an efficient manner. It could also predict the body force when combined with the VFM method. It works well on limited or even missing regions of the training data presented on unstructured mesh or scattered points.

In Sec. \ref{sec:methodology}, the problem set-up and methodology are introduced. The principle and structure of the proposed \textcolor{black}{FCN} and its sub-modules are described in detail here, as well as a brief introduction to other networks for comparison.

In Sec. \ref{sec:experiments}, the results for the flow field reconstruction and force prediction through the FCN model described in Section 2 are presented. Extended experiments are also introduced and analyzed in this section. Concluding remarks are given in Section 4.

\section{\label{sec:methodology}Problem set-up and methodology}
We start with a classical flow problem around a circular cylinder. The unsteady fluid motion is governed by the incompressible Navier-Stokes equations where the density of the fluid is a constant $\rho $ and viscosity of the fluid is a constant $ \mu $. The solid body is denoted by $\Omega _{B}$  bounded by a closed surface $ S_{B} $. 
Given scattered measurements of the snapshot data of the flow field, this work aims to infer the fluid dynamics features, such as the velocity field and the body force. Specifically, this work is devoted to accurately predicting the flow field velocity $ (u_{p}, v_{p}) $ on arbitrary nodes $ \mathbf{x}_{p} = (x_{p}, y_{p}) $, from the observable data $ (u_{r}, v_{r}) $ on a finite number of reference nodes $ \mathbf{N}_{r}$ with coordinates $ \mathbf{x}_{r} = (x_{r}, y_{r}) $.  Moreover, with the vortex force map (VFM), we can calculate the body forces (lift and drag) from the inferred velocity fields.

To solve the aforementioned problem, firstly, we use $\mathbf{\mathscr{M}}$ to represent the model. The model output
\begin{equation}
  \widehat{\hbar}_{p} = \mathscr{M}( \hbar_{r}, \mathbf{x}_r, \mathbf{x}_{p} )
  \\
  \label{eq:model-framework}
\end{equation}
is a function of the model inputs: the observed features $ \hbar_{r} $, the coordinates $ \mathbf{x}_{r} $ of the reference nodes, and the coordinates $ \mathbf{x}_{p} $ of the prediction nodes. The outputs of the model are the predicted features $ \widehat{\hbar}_{p} = ( \widehat{u}_{p}, \widehat{v}_{p}) $ on the target nodes. 

Here the subscript $r$ represents the reference nodes while the subscript $p$ represents the prediction nodes. The symbol with a hat, e.g. $ \widehat{\hbar} $, represents the predicted features, while the symbol without a hat ($ \hbar $) represents the ground truth features. Here in this paper, we presume the ground truth features are the data computed from CFD. In this paper, we deal with two-dimensional (2D) flow field completion cases, where the model is defined as $\mathscr{M}_{FCN}$. This 2D model could easily be extend to a three-dimensional (3D) model by extending the 2D N-S loss function Eq. (\ref{eq:Loss-NS}) to 3D and changing the coordinates of relevant nodes ($ X_{p}, X_{adj}, x_{1}...x_{6} $ in Fig~\ref{fig:framework} (d)) to 3D coordinates.
Part of the data are used to train the model and the rest are used to test and evaluate the model. The details of  sampling data set are described in Sec.~\ref{sec:sec2.3}.

After obtaining the velocity fields from the incomplete measurements through the aforementioned model $\mathscr{M}_{FCN}$, we recall the VFM method \citep{li_zhao_graham_2020} to extract the lift and drag coefficients  on the circular cylinder 
\begin{equation}
\left \{ 
\begin{array}{c}
C_L= \frac{2}{V_{\infty }^{2}d} \iint_{\Omega }\overrightarrow{\Lambda }_{L}\bullet 
(u,v)\omega _{z}d\Omega + 
\\
\frac{2}{V_{\infty}Re}( \oint_{l_{B}}\omega _{z}d\phi _{L} +  \oint_{l_{B}}\omega _{z}\overrightarrow{k_L}\bullet \overrightarrow{dl})\\ 
C_D= \frac{2}{V_{\infty }^{2}d} \iint_{\Omega }\overrightarrow{\Lambda }_{D}\bullet (u,v)\omega _{z}d\Omega + 
\\
\frac{2}{V_{\infty}Re}(\oint_{l_{B}}\omega _{z}d\phi _{D} +  \oint_{l_{B}}\omega _{z}\overrightarrow{k_D}\bullet \overrightarrow{dl})\\
\end{array}
\right.
\label{eq:Metric-CLCD}
\end{equation}
where $V_\infty$ is the free stream velocity, $d$ is the diameter of the circular cylinder, and $\omega _{z} = v_{x}-u_{y} $ is the vorticity. $\overrightarrow{k_L}$ and  $\overrightarrow{k_D}$ are the unit vector in the lift and drag directions respectively. $Re = \frac{\rho{V}_{\infty}d}{\mu}$ is the Reynolds number. The vortex force vectors are defined as
\begin{equation}
\left \{ 
\begin{array}{cc}
     \overrightarrow{\Lambda _{L}}=\left( \frac{\left( x^{2}-y^{2}\right) d^{2}}{4%
\left( x^{2}+y^{2}\right) ^{2}},\frac{xyd^{2}}{2\left( x^{2}+y^{2}\right)
^{2}}\right)\\
     \overrightarrow{\Lambda _{D}}=\left( -\frac{xyd^{2}}{2\left(
x^{2}+y^{2}\right) ^{2}},\frac{\left( x^{2}-y^{2}\right) d^{2}}{4\left(
x^{2}+y^{2}\right) ^{2}}\right)
\end{array} \text{,}
\right.
\label{eq-Lamda-cylinder}
\end{equation}%
and the hypothetical potential are defined as 
\begin{equation}
\phi _{L}=\frac{yd^{2}}{4(x^{2}+y^{2})},\phi _{D}=\frac{xd^{2}}{4(x^{2}+y^{2})}%
\text{.}  \label{eq-phi-cylinder}
\end{equation}

\subsection{\label{sec:sec2.1}The \textcolor{black}{FCN}}

\begin{figure*}
  \centerline{
    \includegraphics[width=0.9\textwidth]{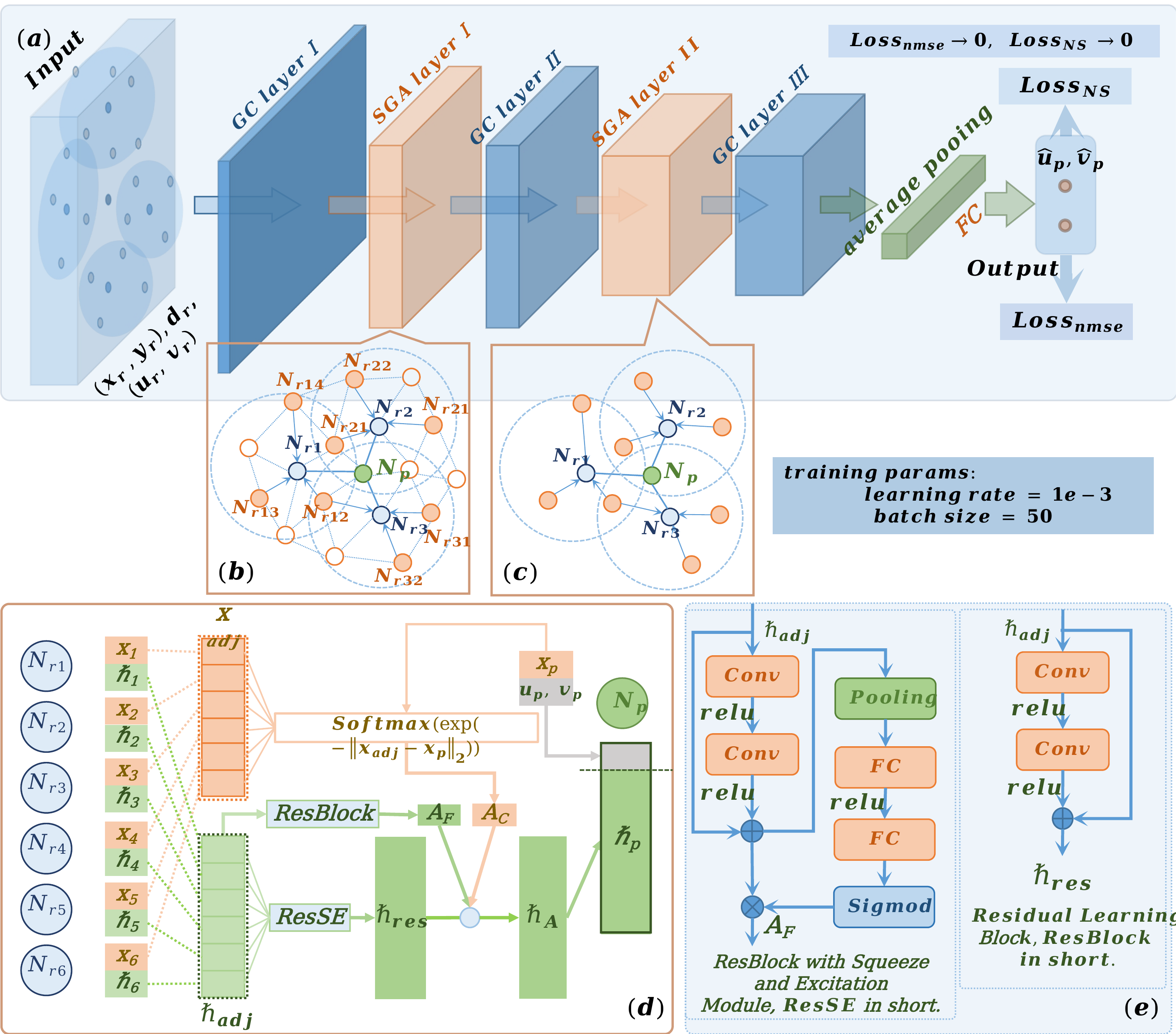}}
  \caption{The structure of the \textcolor{black}{FCN}. \textbf{(a)} The framework of the FCN, which consists of an input and an output layer, three graph convolution layers (GC layers I, II, and III), and two spatial gradient attention layers (SGA layers I and II). 
    \textbf{(b)} Visual illustration of the SGA layer I. The reference nodes ( $N_{rj1},N_{rj2},...,N_{rj6}$ shown as solid orange dots) are used to refer the intermediate nodes ($N_{rj}$ shown as solid blue dots).  \textbf{(c)} Visual illustration of SGA layer II in \textbf{(a)}. Six intermediate nodes ($N_{r1}, N_{r2},...N_{r6}$) are used to refer the prediction node ($N_{p}$ shown as solid green dot). Note that, for illustration purpose only, the structure of the GraphSAGE depicted in this figure comprises 3 reference points in each layer.
    \textbf{(d)} The main framework the SGA layers, which shows how to refer the features on the prediction node $N_p$ from its 6 neighbourhood refernce nodes ($N_{r1}, N_{r2},...N_{r6}$). Here, $ X_p $ and $ \hbar_x $ are the coordinates and the features of node $ p $, respectively. $ A_F $ is features attention, while $ A_C $ is coordinates attention. 'ResBlock' represents the Residual Learning Block, and 'ResSE' represents the ResBlock with Squeeze and Excitation Module. 
    \textbf{(e)}. The details of the ResBlock and ResSE. 'Conv' denotes the one dimensional convolution layer, 'Pooling' denotes the Global average pooling layer, and 'FC' denotes the full connected layer. 
    \label{fig:framework}}
\end{figure*}

The main framework of the proposed FCN is shown in Fig.~\ref{fig:framework}. \textcolor{black}{The FCN consists of three graph convolution (GC) modules (GC layers I, II, III) and two spatial gradient attention (SGA) modules (SGA layers I, II), as shown in Fig~\ref{fig:framework} (a). Each GC layer contains one simple neuron layer, and each SGA layer is a multi-layer perception (MLP) containing 6 simple neuron layers. Thus, the total number of hidden layers is 15, each of them containing 64 neurons as a general treatment to meet the requirements of the model performance~\citep{DNN_DeepVIV_Raissi_2019}.
 The activation functions between different hidden layers are ReLU Activation\citep{Relu2011} (Torch.nn.Relu in the Pytorch deep learning framework). There are no output functions in our model.}
The GC module is mainly used to learn the node features. The details of the structure of each GC module will be introduced in Sec.~\ref{sec:sec2.1.1}. \textcolor{black}{For a more accurate aggregation process, in other words, learning the flow features on the targeting nodes from the neighbor nodes more accurately, the SGA module is extended from the aggregation module in GraphSAGE~\citep{GNN_GraphSAGE_William_2017}. More details could be found in Sec.~\ref{sec:sec2.1.2}. The N-S loss function in Eq.~(\ref{eq:Loss-NS}) enables the SGA modules to be capable of learning the gradient characteristics in line with the physical laws described by the N-S equations~(\ref{eq:NS-equations}). The first and second SGA layers are designed to learn the first-, and second-order partial derivatives in line with the N-S equations respectively. Three GC layers are then designed accordingly before and after the SGA layers to fuse the flow field feature and the learned spatial gradient information. We choose the number of three GC layers, rather than more GC layers, to suppress the over-smoothing phenomenon ~\citep{Graph-OverSmooth-3,Graph-OverSmooth-4,Graph-OverSmooth-5} for a more accurate model.}

\subsubsection{\label{sec:sec2.1.1}The Graph Convolution (GC) Module}
The GC module is introduced from GraphSAGE~\citep{GNN_GraphSAGE_William_2017}. The role of the GC module is to learn the topological structure and the vertex features, in other words, the embedding representation of vertices. \citet{GNN_GraphSAGE_William_2017} proposed a general, inductive framework that leverages node feature information (e.g., text attributes) to efficiently generate node embedding for the previously unseen data - GraphSAGE.
The three GC modules we used is shown in fig.\ref{fig:framework} (a). 

\subsubsection{\label{sec:sec2.1.2}The Spatial Gradient Attention Module}
In the SGA layers, the spatial-based graph convolution network is introduced, where the GraphSAGE (Graph SAmple and aggreGatE) proposed by \citet{GNN_GraphSAGE_William_2017} is adapted here to deal with spatial-based graph. We have made two main aspects of renovations on the original GraphSAGE. The first one is to replace the feature distance layer with the gradient feature attention layer ($ A_F $ in Fig. \ref{fig:framework} (d)) in the updated GraphSAGE model. And the second one is to introduce the gradient coordinate attention layer ($ A_C $ in Fig. \ref{fig:framework} (d)), which is based on spatial coordinate. 
The two SGA modules (SGA I and SAG II) are shown in fig.\ref{fig:framework} (b) and (c). Figure \ref{fig:framework} (b) shows how to refer the data on the target node $ N_{ri} $ from its 6 neighbour nodes $ N_{ri1}, N_{ri2}, ..., N_{ri6}$. For demonstration purposes, only 3 neighbour nodes are depicted in the figure. Similarly, fig.\ref{fig:framework} (c) shows how to refer the data on the target node $ N_{p} $ from its 6 neighbour nodes $ N_{r1}, N_{r2}, ..., N_{r6}$ and only 3 neighbour nodes are depicted.

The SGA module is proposed here to calculate the spatial gradient and aggregate it into the node features. \textcolor{black}{As the NS equations in Eq.~(\ref{eq:NS-equations}) contain the second-order partial derivatives, two SGA layers are introduced in our model, to learn to compute the first and second partial derivatives.} In order to aggregate node features from multi local neighborhood nodes, We modify the input channel of the attention layers ($ A_F $) from original GraphSAGE frameworks. As shown in figures \ref{fig:framework} (d) and (e), the gradient attention layer contains a ResBlock module and a ResSE module. The ResSE is the SE-ResNet Module proposed in \citet{DL_SENet,DL_ResNet} and the ResBlock is the  Residual module introduced by \citet{DL_ResNet}. 

\subsection{\label{sec:sec2.2}Other Models}
As for comparisons to our FCN, \emph{the DNN-based model} and \emph{the CNN-based model}, commonly used in fluid dynamics, are also introduced here.

\begin{figure}
  \centerline{
    \includegraphics[width=0.5\textwidth]{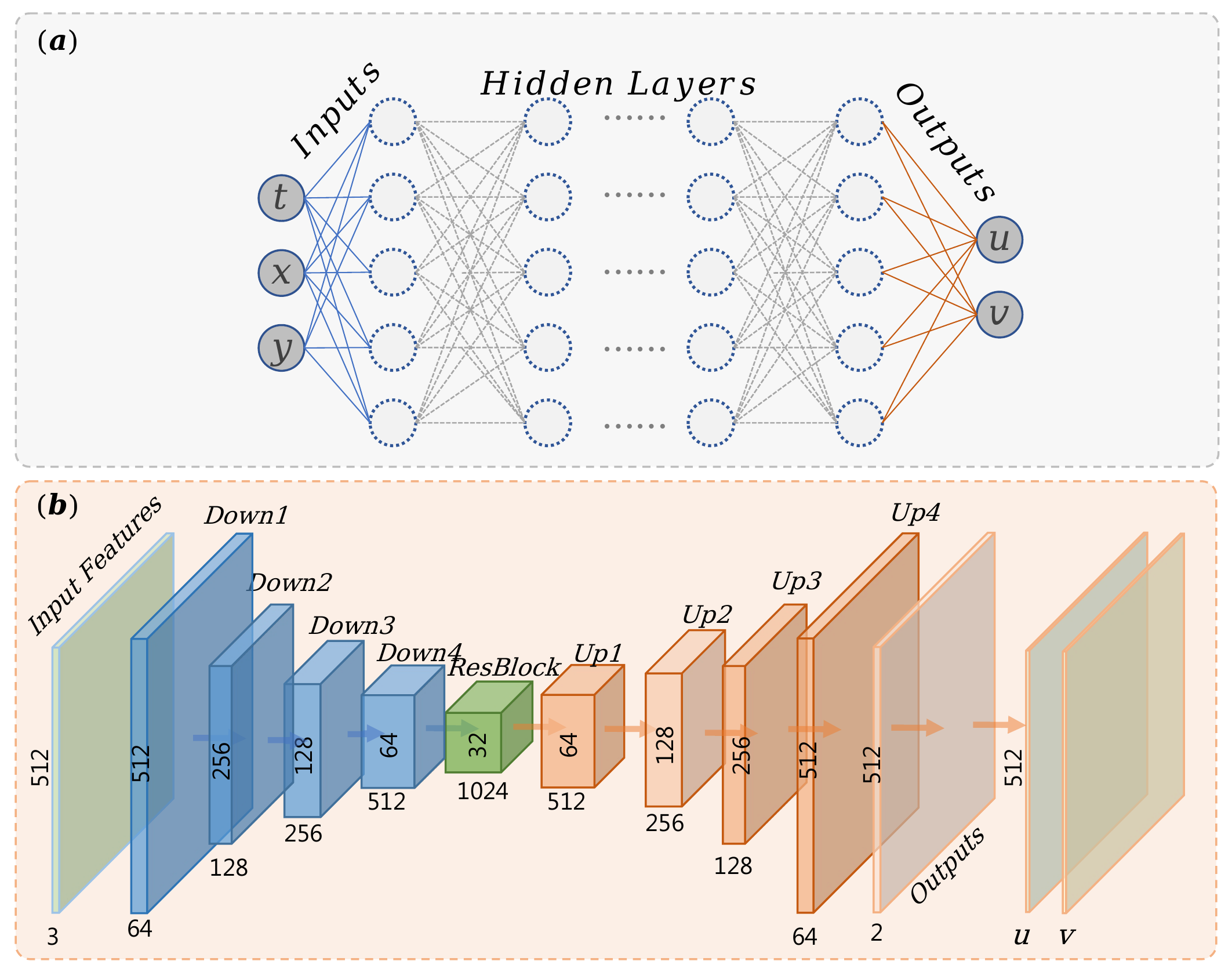}
    }
  \caption{The main framework of the other models. 
  (a) is the DNN based model $ M_{DNN} $.  
   The DNN  consists of 10 hidden layers and 64 weight neurons and 1 bias neuron per hidden layer. The DNN takes the input variables t, x and y and outputs u and v.
  (b) is the CNN based model $ M_{CNN}$.}
  \label{fig:model.othermodels}
\end{figure}

\paragraph{\emph{The DNN-based model}} $\widehat{\hbar}_{p} = \mathit{\mathscr{M}_\mathit{DNN}}( t_{p}, \mathbf{x}_{p} )$ relates the input data (time $ t $ and the coordinates $ x $) to the predicted features $\widehat{\hbar}_{p}$ on the target nodes. The main framework of $ \mathscr{M}_\mathit{DNN} $ is shown in Fig.~\ref{fig:model.othermodels} (a). The DNN-based model contains 10 hidden layers, each hidden layer consists of 64 weight neurons and 1 bias neuron. The normal square error loss equation~(\ref{eq:Loss-nmse}) and the Navier-Stokes loss functions are used in the training procedure for our DNN-based model, following the work by~\citet{DNN_DeepVIV_Raissi_2019}.

\paragraph{ \emph{The CNN-based model}}  $\widehat{\hbar}_{p} = \mathit{\mathscr{M}_{CNN}}( \hbar_{r}, m )$ shown in Fig.~\ref{fig:model.othermodels} (b) is similar to the deep learning model proposed by \citet{DL-Impainting-2017}. The input data of the model are the given features $ \hbar_{r} $ of the given nodes and the mask $ m $, and the output data is the features $\widehat{\hbar}_{p}$ on the targeting nodes. Here, the mask $ m $ is used to distinguish the reference nodes from the target nodes, $ m=0 $ refers to the reference node, while $ m =1 $ refers to the target node. This model uses a SegNet-like framework~\citep{DL_SegNet_2017} consisting of encoding layers and decoding layers. Each of the encoding layers contains 4 downsample blocks {Down1, Down2, Down3, Down4} as shown in Fig.~\ref{fig:model.othermodels}, and each of the decoding layers also contains 4 upsample blocks ( Up1, Up2, Up3, Up4) as shown in Fig.~\ref{fig:model.othermodels}. A ResBlock~\citep{DL_ResNet} is used to connect the encoder and the decoder in the CNN-based model. The Mean Square Error loss, same as the loss functions in SegNet\citep{DL_SegNet_2017}, is used here for the training of the CNN-based model.

\subsection{\label{sec:sec2.3}Dataset, Loss Functions, and Metrics}
The flow field data around a circular cylinder obtained by CFD are used as the dataset in this work. \textcolor{black}{For CFD simulations, the Navier–Stokes equations in unsteady laminar flow are solved numerically using the same method as used by Li et al. (2021)\citep{Lietal2021}. It consists of using the commercial code Fluent with the options of a second-order upwind SIMPLE (semi-implicit method for pressure-linked equations) pressure–velocity coupling method. The computational domain is 36 diameters in the inflow
direction and 21 diameters in the direction perpendicular to the inflow. For the cases of $ Re = 100 $, $ Re = 500 $, and $ Re = 1000 $, three meshes with 41,573, 64,190, and 80,197 grid points are used, respectively. For all the cases, 20 mesh layers inside the laminar boundary layer are guaranteed.  The flow is impulsively started from an initially uniform flow. 
The non dimensional time $\tau = {t}{V}_{\infty}/d$ is defined as the number of chord length the uniform flow travels. For the cases of $ Re = 100 $, $ Re = 500 $, and $ Re = 1000 $, a total time of $\tau = 150, 50, and 50$ are simulated. For every $\Delta{\tau} = 1$, the flow field data, including the coordinates, the velocity, and the vorticity, are saved to form the dataset. Thus, the dataset contains a total points of $N_{T}{(Re = 100) = 41,573\times 150}$ for the case of $Re = 100$, it consists of $N_{T}{(Re = 500) = 64,190\times 50}$ for the case of $Re = 500$, and it include $N_{T}{(Re = 1000) = 80,197\times 50}$ for the case of $Re = 1000$. }

\textcolor{black}{The whole dataset are divided into three parts: the training set for the training process of our models, the validation set for finding the proper hyper-parameters, and the test set for assessing the performance of the model. To ensure the reliability of the model in the general cases, there is no overlap between the training and test datasets.}
\textcolor{black}{The selection of various datasets are done based on a random process~\citep{POURZANGBAR2017107}, i.e., 10-fold cross validation strategy here. This random process may lead to spatial imbalance in the dataset. In order to avoid the data imbalance, several data augment strategies are introduced into the model training process, which will be detailed in Sec.~\ref{sec:sec3.1.1}.}
\textcolor{black}{Note that the CFD method utilized in this work is purely for providing data and validation purposes. It is the intent for future work that we implement the proposed FCN model for prediction of an experimental field dataset. We should mention that the entire methodology of this work does not involve a pressure field, making its application to experimental data, such as PIV, more instinctive. The numerical method used in this work has been well validated in the previous work by Li et al. \citep{LIWU2018,Lietal2021}, thus the details of the validation are not given here.} 

\textcolor{black}{We carried out three sets of experiments with different size of training dataset: 10\%, 30\%, and 50\%, respectively. The according validation dataset and testing dataset are shown in table~\ref{tab:dataset}. As in the literature, some existing models \citep{Ali2021,DNN_DeepVIV_Raissi_2019} use >80\% of data for training, and the rest are split equally as the validation dataset and testing dataset. However, the large proportion of the training dataset sets an inevitable limitation for the generalization ability of existing models. Thus, in this work, we tried to train our model with a smaller proportion (10\%, 30\%, and 50\%) of training datasets. Moreover, a larger proportion of the testing set than the validation set is chosen to show the good performance of our proposed model.}
\textcolor{black}{The focus of this paper is to design and validate this flow completion network. To show the reliability of the model, three different experiments has been carried out to train three models for three different Reynolds numbers ($ Re = 100 $,$ Re = 500 $,$ Re = 1000 $), separately. An updated model capable of dealing with multi-Reynolds numbers with rigorous uncertainty and reliability assessment is worth to be explored in future work.}
To evaluate the performance of all the models, all the metrics for the test dataset are computed to assess the capability of the model trained by different training data sets. \textcolor{black}{The $ u $ and $ v $ obtained by CFD are used as labels in our dataset.}


\begin{table}
  \centering
    \textcolor{black}{
  \begin{tabular}{lccc}
  \\\hline
    Train dataSet: & 10\%$N_T$ & 30\%$N_T$ & 50\%$N_T$ \\\hline
    Validate dataset: & 30\%$N_T$  &  30\%$N_T$    & 20\%$N_T$   \\
    Test dataset:   & 60\%$N_T$  & 40\%$N_T$     & 30\%$N_T$   \\\hline
  \end{tabular}
    }
  \caption{\textcolor{black}{The detail of the dataset. As comparison, we have done the same experiments on different training dataset: 10\%, 30\%, and 50\%, respectively. Here $N_{T}= 41,573\times 150$ for the case of $Re = 100$, $N_{T} = 64,190\times 50$ for the case of $Re = 500$, and $N_{T}{ = 80,197\times 50}$ for the case of $Re = 1000$.}}
  {\label{tab:dataset}}
\end{table}

We use the normal mean-square error and the absolute error as the metrics to evaluate the performance of the model. The proportion of the training set is also an important index to measure the generalization ability of the model. If the results of the model are the same, the fewer data used in the model training process, the stronger the generalization ability of the model. $ \hbar_{p} = ( u_{p}, v_{p} ) $ is the $ u $ and $ v $ of the CFD result, $ \widehat{u}_{p}, \widehat{v}_{p} $ is predicted $ u $ and $ v $ for the target nodes by the model $M$. \textcolor{black}{The equation of the normal mean-square error $ E_{nmse}$ for both $u$ and $v$ are as follows:}
\begin{equation}
  \begin{array}{c}
    E_{nmse} (u) = \frac{\left\lVert \widehat{u}_p - u_p\right\rVert_2}{\left\lVert u_p \right\rVert_2},    E_{nmse} (v) = \frac{\left\lVert \widehat{v}_p - v_p\right\rVert_2}{\left\lVert v_p \right\rVert_2},
  \end{array}%
  \label{eq:Metric-nmse}
\end{equation}\begin{equation}
  \begin{array}{c}
    NMSE (u,v) =  \sqrt{((E_{nmse}(u))^2  +(E_{nmse}(v))^2)/2},
  \end{array}%
  \label{eq:NMSE-total-uv}
\end{equation}

\textcolor{black}{To assess the generality of the proposed model, in addition to the NMSE of the flow velocity, the relative error of the force coefficients extracted from the predicted $ u $ and $ v $ and the coefficients computed by CFD} 
\begin{equation}
  \begin{array}{c}
    E_{r} ( C_L ) = \frac{| \widehat{C_L} -  C_L |}{| \Delta {C_L} |}, E_{r} ( C_D ) = \frac{| \widehat{C_D} -  C_D |}{| \Delta {C_D} |}
    \label{eq:Metrics-CLCD}
  \end{array}
\end{equation}
are also utilized in evaluating the performance of our model.

\textcolor{black}{Other statistical error parameters are also defined, including the Correlation Coefficient (CC)}
\begin{equation}
  \begin{array}{c}
    E_{CC}(u) = \frac{\left\lvert (u_p-\overline{u_p})(\widehat{u_p}-\overline{\widehat{u_p}})\right\rvert }{\left\lVert u_p-\overline{u_p} \right\rVert_2 \left\lVert \widehat{u_p}-\overline{\widehat{u_p}} \right\rVert_2} , E_{CC}(v) = \frac{\left\lvert (v_p-\overline{v_p})(\widehat{v_p}-\overline{\widehat{v_p}})\right\rvert }{\left\lVert v_p-\overline{v_p} \right\rVert_2 \left\lVert \widehat{v_p}-\overline{\widehat{v_p}} \right\rVert_2} \\
    \\
    CC(u,v) = \sqrt{((E_{CC}(u))^2  +(E_{CC}(v))^2)/2},

    \label{eq:Metrics-CC}
  \end{array}
\end{equation}
where the subscript $p$ indicates the nodes of prediction (or the target nodes), the characters without hat (e.g. ${u_p}$ and ${v_p}$) mean the ground truth and the characters with a hat (e.g. $\widehat{u_p}$ and $\widehat{v_p}$) mean the predicted value. The averaging of the variables is indicated by an overline (e.g.$\overline{u_p}$ and $\overline{v_p}$).

\textcolor{black}{The Bias is defined as}
\begin{equation}
  \begin{array}{c}
    E_{bias}(\frac{u}{V_{\infty}}) = \overline{\frac{\widehat{u_p}}{V_{\infty}}- \frac{u_p}{V_{\infty}}}, E_{bias}(\frac{v}{V_{\infty}}) = \overline{\frac{\widehat{v_p}}{V_{\infty}} - \frac{v_p}{V_{\infty}}}. 
  \end{array}%
  \label{eq:Metrics-BIAS}
\end{equation}

The motion of fluids is expressed by conservation laws for mass, momentum and energy. The equation for mass is known as the continuity equation while the equation for momentum is called equation of motion that is an expression of Newton's law. If viscous fluid and inviscid fluid are considered in these equations, they are known as the Navier-Stockes and Euler equations, respectively.

Two loss functions are introduced to the training procedure for our model. \textcolor{black}{One is  the loss function $Loss_{Grad-NS}$ specialized here to include the the partial derivatives in the N-S equations}
\begin{equation}
 \left \{
  \begin{array}{c}
    u_{t}+uu_{x}+vu_{y}=-p_{x}+Re^{-1}\left( u_{xx}+u_{yy}\right) \\
    v_{t}+uv_{x}+vv_{y}=-p_{y}+Re^{-1}\left( v_{xx}+v_{yy}\right) \\
    u_{x}+v_{y}=0%
    \label{eq:NS-equations}
  \end{array}\right.
  \text{,}%
\end{equation}
\textcolor{black}{and another one is the conventional normal mean square error loss function $Loss_{NMSE}$. The $Loss_{Grad-NS}$ is designed to guide the model to learn the gradient information in the flow field and is defined as}
\begin{equation}
  \begin{array}{c}
  Loss_{Grad-NS}(\widehat{\hbar}_{p}, \hbar_{p}, \mathbf{x}_{p}, Re) = 
  || \widehat{u}_{x} - u_{x}||_2  + 
  || \widehat{u}_{y} - u_{y}||_2  + 
  \\
  Re^{-1}||\widehat{u}_{xx} - u_{xx}||_2 + 
  || \widehat{u}_{yy}  - u_{yy} ||_2 +
  || \widehat{v}_{x} - v_{x} ||_2 +  
  \\
  || \widehat{v}_{y} - v_{y} ||_2 + 
  Re^{-1}|| \widehat{v}_{xx} v_{xx}||_2 + 
  || \widehat{v}_{yy} - v_{yy} ||_2 + 
  \\
  || \widehat{u}_{x}+\widehat{v}_{y} ||_2 \text{.}%
  \label{eq:Loss-NS}
  \end{array}
\end{equation} 
The $Loss_{NMSE}$ is defined as
\begin{equation}
  Loss_{NMSE}(\widehat{\hbar}_{p}, \hbar_{p}) = \frac{|| u_{p} - \widehat{u}_{p} ||_2}{|| u_{p} ||_2 + \epsilon } + \frac{|| v_{p} - \widehat{v}_{p} ||_2}{|| v_{p} ||_2 + \epsilon}
  \text{.}%
  \label{eq:Loss-nmse}
\end{equation}
Here, $ \epsilon = 1e-5  $ is added to its denominator to avoid numerical errors. To train the model $\mathit{\mathscr{M}_{FCN}}$, we aim to minimize the difference between the predicted $ \widehat{\hbar}_{p} $ and the $ \hbar_{p} $ from the CFD result by introducing the loss functions $ Loss_{NS}(\widehat{\hbar}_{p}, \hbar_{p}, \mathbf{x}_{p}, Re) $ and  $ Loss_{nmse}(\widehat{\hbar}_{p}, \hbar_{p})$ .

\section{\label{sec:experiments}Experiments and Results}
\subsection{\label{sec:sec3.1}Training}
The training dataset mentioned in Sec. \ref{sec:sec2.3} is used to train different models in our experiments. \textcolor{black}{In order to obtain widely applicable models, dimensionless parameters (such as $u/V_{\infty},v/V_{\infty},2a\omega/V_{\infty}$, and $\tau = tV_{\infty}/(2a)$), instead of dimensional parameters (such as $u,v,\omega$, and $t$) are used in this work.} The Flow Completion Network converges gradually after about 40 epochs of training. The CNN-based model and DNN-based model, used as comparisons here, converge after about 61 and 67 epochs of training, respectively.

\subsubsection{\label{sec:sec3.1.1}The Data Augmentation Strategies}

In order to improve the generalization of the model, several data augmentation strategies are used during the training procedures. The balance-weight sampling method, gaussian noise method are introduced into our training procedures.

\textit{The Balanced Weight Sampling Method.} To guide the model to better learn the flow features close to the wall area and predict the flow more accurately, we add the sampling weight $ w = 1/(\sqrt{x^2+y^2} - 0.5*r ) $ to the nodes close to the wall. The nodes closer to the wall have a larger sampling weight, while the nodes farther away from the wall have a smaller sampling weight.

\textit{The Gaussian Nose Method.} To enhance the fitness of our models, the gaussian noise is added to the original features. The Gaussian noise with zero mean value and $ 1.0 $ standard deviation $ n_{f} \sim N(0,1.0) $ is added to the original features $ \hbar_{r} $ during training procedure.

\subsubsection{\label{sec:sec3.1.2}The Hyper-parameters in the Training}

The hyper-parameters for the training procedure of our model are shown in the following list:
\begin{itemize}
  \item Optimizer \textcolor{black}{(Training algorithm): stochastic gradient descent (SGD) algorithm};
  \item Momentum: 0.9;      
  \item Learning rate: 10e-3;
  \item Batch size: 32;
  \item Training epochs: 100;
  \item Dropout Rate: 0.5.
\end{itemize}

\begin{figure}
  \centerline{
    \includegraphics[width=0.5\textwidth]{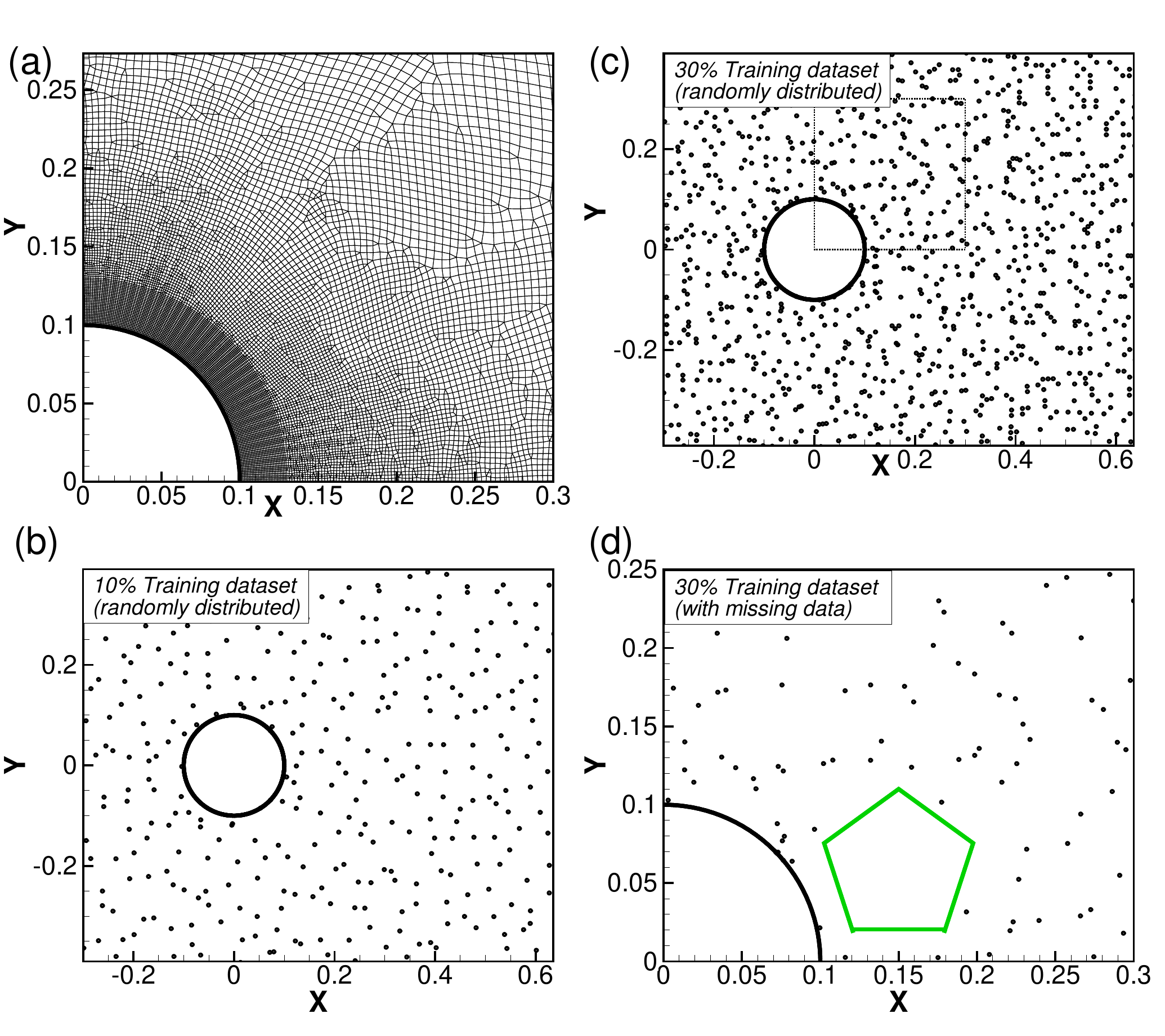}}
  \caption{\textcolor{black}{For the case of Re = 1000.} (a) The schematic of the unstructured mesh grid used in the CFD. (b) The 10\% randomly distributed training dataset nodes in the flow field. (c) The 30\% randomly distributed training dataset nodes in the flow field without unobserved regions. (c) The 30\% randomly distributed training dataset nodes in the flow field with an unobserved pentagon region in the wake area which is shown in green.}
  \label{fig:mesh_training_dataset}
\end{figure}

\textcolor{black}{The best learning rate and batch size are determined by the grid search strategy.} The schematics of the unstructured mesh grid used in the CFD simulation and the training data set are shown in Fig.~\ref{fig:mesh_training_dataset}. As stated in Sec. 2.3, 10\%, 30\%, and 50\% uniform randomly distributed scatters are subtracted from the total dataset as training datasets. In Fig.~\ref{fig:mesh_training_dataset}, we only show the 10\% and 30\% training dataset for simplicity. Meanwhile, the training data set with and without unobserved regions are also tested.

\subsection{Results}

The experimental results on the dataset described in Sec.~\ref{sec:sec2.3} are presented in this section. The proposed FCN model is compared with the traditional CNN and DNN-based models in the same dataset. 
We also test the generalization ability of all the models on the dataset. 

\begin{figure*}
  \centerline{
    \includegraphics[width=0.87\textwidth]{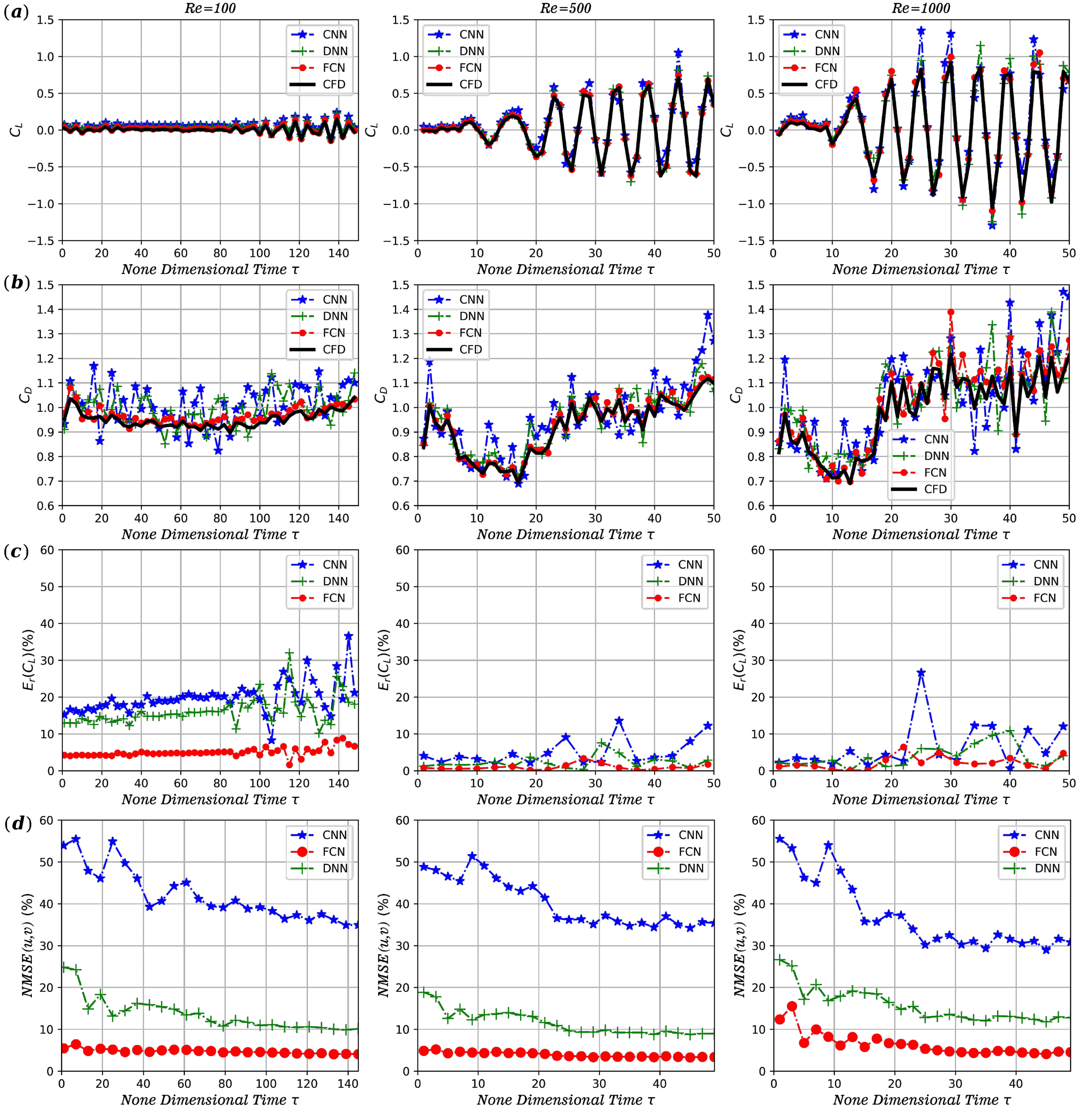}
    }
  \caption{\textcolor{black}{(a) The comparison of the lift coefficient curve predicted by $\mathit{\mathscr{M}_{FCN}}$, $\mathit{\mathscr{M}_{CNN}}$, $\mathit{\mathscr{M}_{DNN}}$, and computed from the CFD data at three different Reynold numbers: $ Re = 100 $, $ Re = 500 $, and $ Re = 1000 $.
  (b) The comparison of the drag coefficient curve predicted by $\mathit{\mathscr{M}_{FCN}}$, $\mathit{\mathscr{M}_{CNN}}$, $\mathit{\mathscr{M}_{DNN}}$, and computed from the CFD data at three different Reynold numbers: $ Re = 100 $, $ Re = 500 $, and $ Re = 1000 $.
  (c) The relative error of the
lift coefficients obtained from different models ($\mathit{\mathscr{M}_{FCN}}$, $\mathit{\mathscr{M}_{CNN}}$ and $\mathit{\mathscr{M}_{DNN}}$) for three different Reynold numbers cases: $ Re = 100 $, $ Re = 500 $, and $ Re = 1000 $.   (d) The normal mean square error (NMSE) of the flow velocity $NMSE(u,v)$ for three different models ($\mathit{\mathscr{M}_{FCN}}$, $\mathit{\mathscr{M}_{CNN}}$ and $\mathit{\mathscr{M}_{DNN}}$).}}
  \label{fig:result.nmse_clcd}
\end{figure*}

The experiment results of the lift and drag coefficients, against non-dimensional time $\tau = tV_{\infty}/d$, predicted by the $\mathit{\mathscr{M}_{FCN}}$, $\mathit{\mathscr{M}_{CNN}}$, and $\mathit{\mathscr{M}_{DNN}}$ models for different Reynolds numbers are shown in Fig.~\ref{fig:result.nmse_clcd} (a) and (b).
\textcolor{black}{Figure \ref{fig:result.nmse_clcd} (c) shows the relative error of the lift coefficients obtained from different models (CNN, DNN, and FCN) for the three different Reynolds number cases.} The above figures show that our proposed FCN has a better performance in predicting the lift and drag coefficients than the traditional CNN and DNN-based models. 
From Fig.~\ref{fig:result.nmse_clcd} (d), we can see that the normalized mean squared error of our proposed model $\mathit{\mathscr{M}_{FCN}}$ is lower than those of $\mathit{\mathscr{M}_{CNN}}$ and $\mathit{\mathscr{M}_{DNN}}$. Moreover, the performance of the $\mathit{\mathscr{M}_{CNN}}$ and $\mathit{\mathscr{M}_{DNN}}$ varies with non-dimensional time $\tau$ while the proposed $\mathit{\mathscr{M}_{FCN}}$ has a stable performance over all time range. One possible explanation is that our proposed $\mathit{\mathscr{M}_{FCN}}$ could learn physics from the NS loss functions, and utilizes the information on the neighbor nodes, which lead to a better prediction.

\begin{figure*}
  \centerline{
    \includegraphics[width=1.2\textwidth]{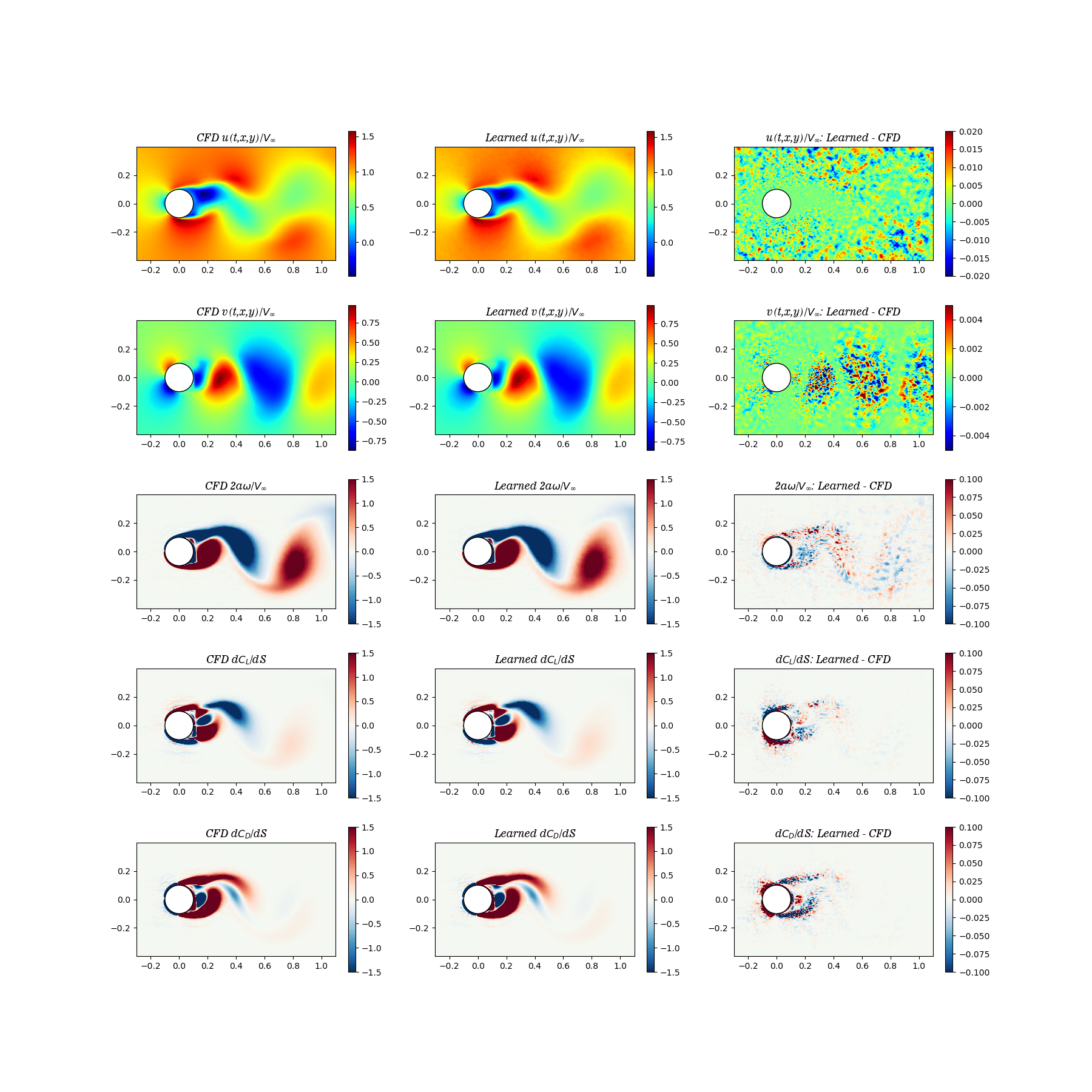}
    }
  \caption{The comparison of the predicted, CFD, and there differences ($Learned - CFD$) on the velocity field ($ u(t,x,y)/V_{\infty}, v(t,x,y)/V_{\infty}$), the vorticity field, and the lift and drag distributions for the case of $Re = 1000$.}
  \label{fig:result.omega.cylinder.1000}
\end{figure*}

The predicted velocity, vorticity, thus lift and drag distribution for the cylinder dataset (Re = 1000) obtained from $\mathit{\mathscr{M}_{FCN}}$ and from CFD at a typical instant are shown in Fig. \ref{fig:result.omega.cylinder.1000}.
The first two lines of Fig. \ref{fig:result.omega.cylinder.1000} show the predicted, CFD, and there differences ($Learned - CFD$) on the velocity field ($ u(t,x,y)/V_{\infty}, v(t,x,y)/V_{\infty}$). The third line of Fig. \ref{fig:result.omega.cylinder.1000} shows the predicted, CFD, and there differences ($Learned - CFD$) on the vorticity field. While the fourth and fifth lines of of Fig. \ref{fig:result.omega.cylinder.1000} show the predicted, CFD, and there differences ($Learned - CFD$) on the lift and drag distributions respectively. A good comparison has been found between our proposed FCN model and the CFD results.

\begin{figure}
  \centerline{
    \includegraphics[width=0.5\textwidth]{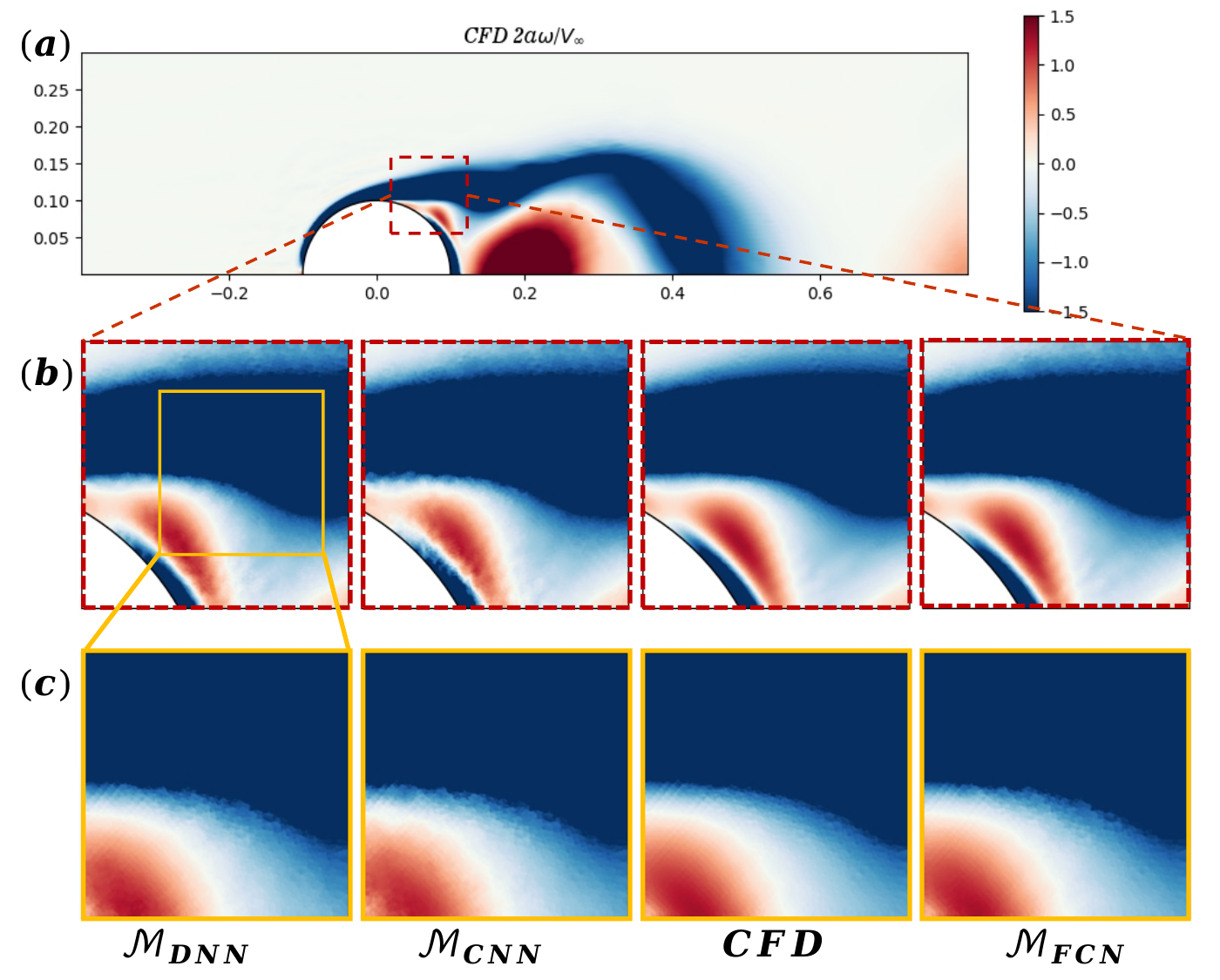}
    }
  \caption{
    The super-resolution flow field comparison of $\mathit{\mathscr{M}_{DNN}}$, $\mathit{\mathscr{M}_{CNN}}$, $ CFD $ and $\mathit{\mathscr{M}_{FCN}}$ for the cylinder dataset (Re = 1000).}
  \label{fig:result.all.models.cylinder.1000}
\end{figure}

In order to demonstrate the effectiveness of our $\mathit{\mathscr{M}_{DNN}}$ model, the flow fields completed by our model, the $\mathit{\mathscr{M}_{CNN}}$ model and the $\mathit{\mathscr{M}_{FCN}}$ model are compared with $ CFD $ results at different scales, shown in Fig.\ref{fig:result.all.models.cylinder.1000}. The vorticity $\omega$ is demonstrated here. From Fig. \ref{fig:result.all.models.cylinder.1000} we can see that the CNN and DNN-based model learned vorticity field is noisy due to a lack of the second-order gradient information during the training procedure. Our proposed FCN defeats the traditional CNN and DNN-based models regarding the flow filed prediction from unstructured data.

\begin{figure}
  \centerline{
    \includegraphics[width= 0.5\textwidth]{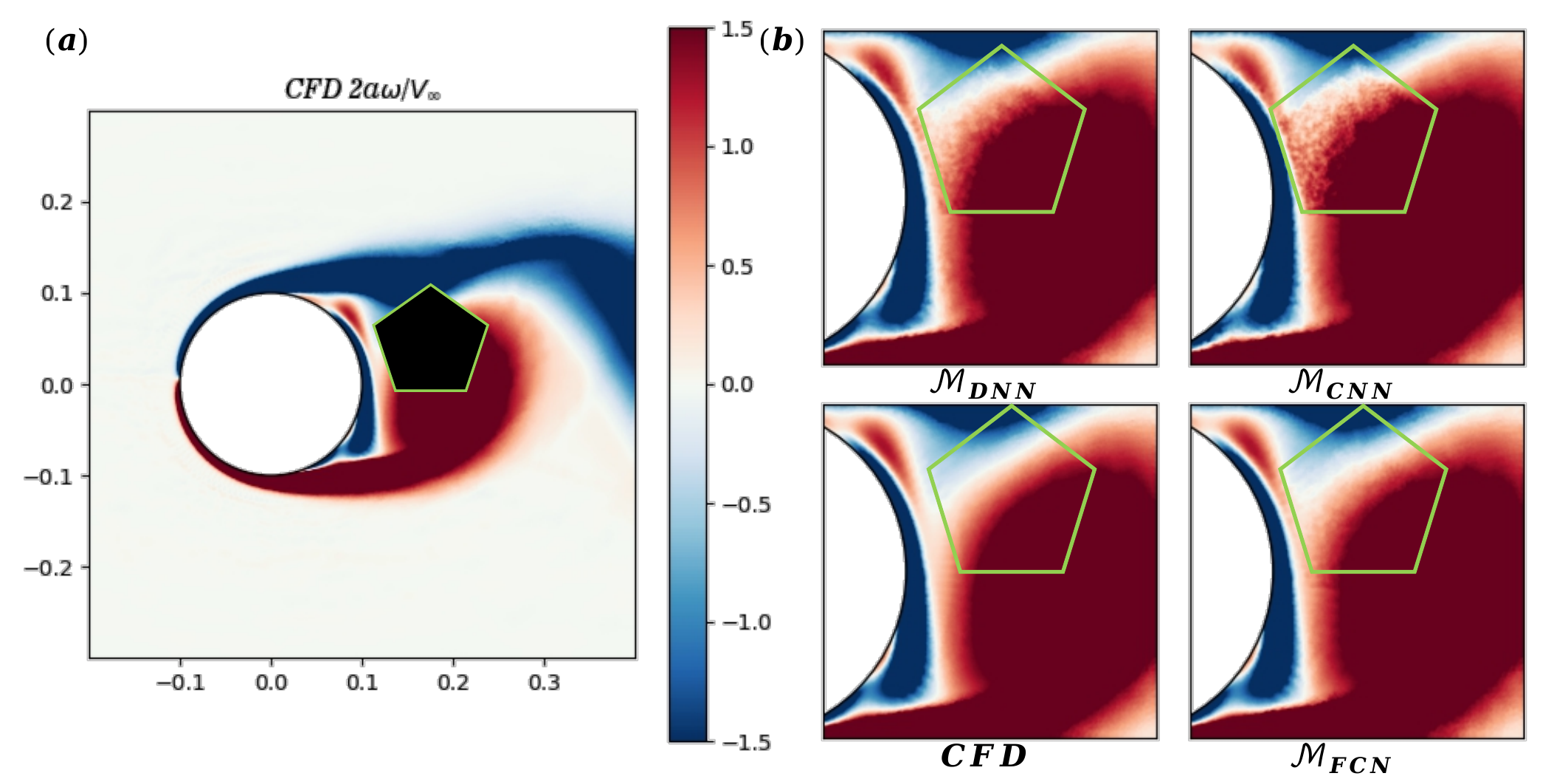}
    }
  \caption{
    The completion flow field for the cylinder dataset (Re = 1000) obtained from $\mathit{\mathscr{M}_{DNN}}$, $\mathit{\mathscr{M}_{CNN}}$, $ CFD $ and $\mathit{\mathscr{M}_{DNN}}$. The black pentagonal area in Fig. (a) is the flow field region that needs to be completed.} 
  \label{fig:result.completion.models.cylinder.1000}
\end{figure}

It is very common to have shadows or unobserved data in the experimental measurements or point-could represented data. The unobserved data may lie in any arbitrary training domains which could contain important information. Here in this work, we use a pentagonal region in the wake of the cylinder as an example. The performance of our proposed $\mathit{\mathscr{M}_{FCN}}$ model compared with  $\mathit{\mathscr{M}_{DNN}}$, and $\mathit{\mathscr{M}_{CNN}}$ model with this missing pentagonal region is also tested. The results are shown in Fig. \ref{fig:result.completion.models.cylinder.1000}, where we could see that the $\mathit{\mathscr{M}_{FCN}}$ completion compares well with the CFD results, while the other two traditional

\begin{figure*}
  \centerline{
    \includegraphics[width=\textwidth]{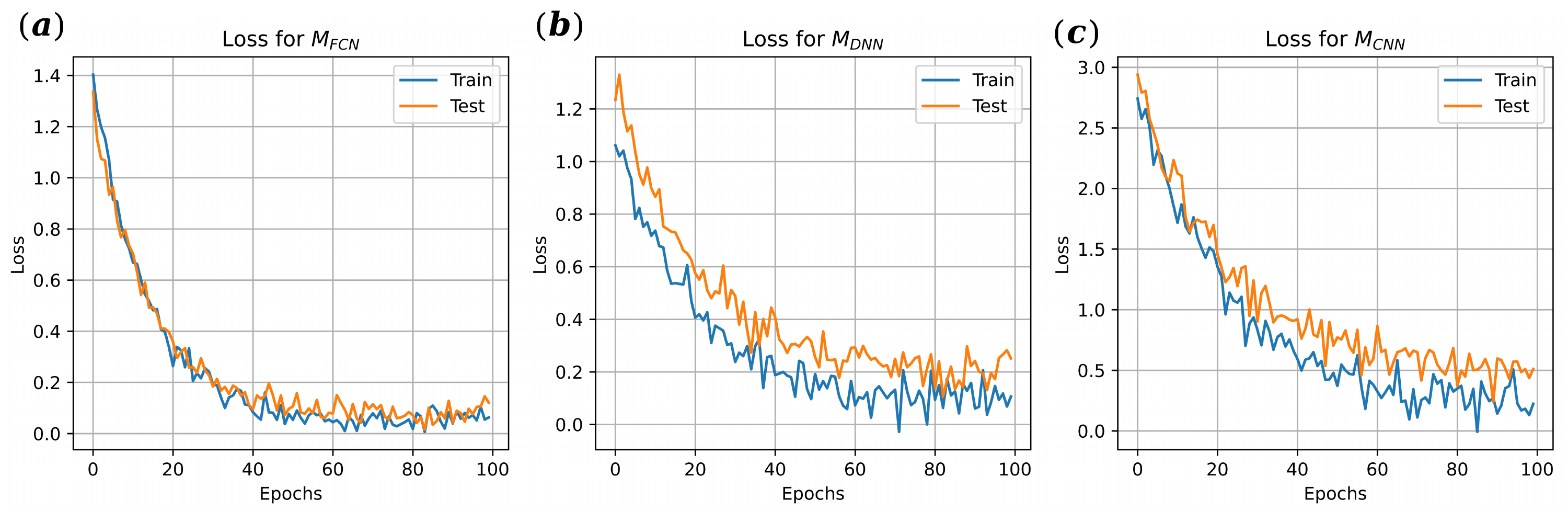}
    }
  \caption{The training loss and testing loss during the training procedure for (a) the proposed FCN model; (b) the CNN-based model; and (c) the DNN-based model.}
  \label{fig:loss}
\end{figure*}

 Fig.~\ref{fig:loss} shows the training loss and testing loss during the training procedure for the $\mathscr{M}_{FCN}$, the $\mathscr{M}_{CNN}$, and the $\mathscr{M}_{CNN}$, respectively. Fig.~\ref{fig:loss} (a) shows that the FCN model converges after about 40 epochs of training. The average training loss converges to 0.0806, and the average testing loss converges to 0.1092 which is very close to the average training loss. The gap between the average testing loss and the average training loss indica almostes that \textcolor{black}{there is almost no overfitting during the training procedure of the proposed FCN model.}
  Similarly, in Fig.~\ref{fig:loss} (b) and (c), the training loss and testing loss are shown to converge after about 61 and 67 epochs of training, respectively. 
  The average training loss and testing loss for the CNN-based model converge to 0.1042 and 0.2324 respectively. The average training loss and testing loss for the DNN-based model converge to 0.0639 and 0.1031 respectively. The convergence rate for the FCN model is the highest among all the models tested here.

\begin{table}
  \centering
  \begin{tabular}{cccc}
  \\\hline
  Partition Ratio & $ \mathscr{M}_{CNN} $(test/train) & $ \mathscr{M}_{DNN}$(test/train) & $ \mathscr{M}_{FCN}$(test/train) \\\hline
  10\% & 37.23\%/19.15\%  & 15.63\%/8.98\%  & 5.86\%/9.79\%  \\
  30\% & 33.09\%/20.85\%  & 10.60\%/12.18\%  & 4.72\%/6.37\%  \\
  50\% & 30.05\%/20.21\%  & 15.22\%/8.46\%  & 3.21\%/6.86\%   \\\hline
  \end{tabular}
  \caption{\textcolor{black}{The metrics ($ NMSE(u,v) $ ) of training dataset and testing dataset for $\mathscr{M}_{CNN}$, $\mathscr{M}_{DNN}$ and $\mathscr{M}_{FCN}$ trained by 10\%, 30\% and 50\% cylinder dataset for the case of $ Re = 1000 $.}}
  \label{tab:results_nmse_re_1000}
\end{table}

\begin{table}
  \centering
  \begin{tabular}{cccc}
    \\\hline
    Partition Ratio & $ \mathscr{M}_{CNN} $(test/train) & $ \mathscr{M}_{DNN}$(test/train) & $ \mathscr{M}_{FCN}$(test/train) \\\hline
    10\% & 40.31\%/20.14\%  & 11.37\%/10.28\%  & 3.84\%/7.18\%  \\
    30\% & 31.05\%/21.27\%  & 12.36\%/10.71\%  & 4.15\%/6.54\%  \\
    50\% & 26.88\%/18.87\%  & 11.98\%/7.78\%  & 4.39\%/4.68\% \\\hline
    \end{tabular}
  \caption{\textcolor{black}{The metrics ($NMSE(u,v)$ ) of training dataset and testing dataset for $\mathscr{M}_{CNN}$, $\mathscr{M}_{DNN}$ and $\mathscr{M}_{FCN}$ trained by 10\%, 30\% and 50\% cylinder dataset for the case of $ Re = 500 $.}}
  \label{tab:results_nmse_re_500}
\end{table}

\begin{table}
  \centering
  \begin{tabular}{cccc}
    \\\hline
    Partition Ratio & $ \mathscr{M}_{CNN} $(test/train) & $ \mathscr{M}_{DNN}$(test/train) & $ \mathscr{M}_{FCN}$(test/train) \\\hline
    10\%            & 42.32\%/22.07\%                   & 13.51\%/10.62\%                    & 4.64\%/7.31\%                 \\
    30\%            & 29.20\%/20.21\%                   & 13.12\%/9.75\%                     & 3.59\%/7.32\%                  \\
    50\%            & 25.45\%/19.96\%                   & 11.92\%/6.85\%                     & 3.82\%/6.38\%                  \\\hline
  \end{tabular}
  \caption{\textcolor{black}{The metrics ($NMSE(u,v)$ ) of training dataset and testing dataset for $\mathscr{M}_{CNN}$, $\mathscr{M}_{DNN}$ and $\mathscr{M}_{FCN}$ trained by 10\%, 30\% and 50\% cylinder dataset for the case of $ Re = 100 $.}}
  \label{tab:results_nmse_re_100}
\end{table}

\textcolor{black}{Tables~\ref{tab:results_nmse_re_1000},~\ref{tab:results_nmse_re_500} and ~\ref{tab:results_nmse_re_100} list the metrics of training dataset and testing dataset for $\mathscr{M}_{CNN}$, $\mathscr{M}_{DNN}$ and $\mathscr{M}_{FCN}$ trained by 10\%, 30\% and 50\% of the total dataset at three different Reynolds numbers: $ Re = 1000 $, $ Re = 500 $, and $ Re = 100 $, respectively.}

\textcolor{black}{In table \ref{tab:results_nmse_re_1000}, we can see that for the case of partition ratio euqals 10\% and $Re = 1000$, the training dataset metric $NMSE(u,v)$ for $\mathscr{M}_{CNN}$ is 37.23\%, higher than its testing dataset metric 19.15\%. Similarly, the testing dadaset metric $NMSE(u,v) = 15.63\%$ for $\mathscr{M}_{DNN}$ is slightly higher than the training dataset metric $NMSE(u,v) = 8.98\%$. While, for our  $\mathscr{M}_{FCN}$ model, the testing dadaset metric $NMSE(u,v) = 5.86\%$ is lower than its training dataset metric $NMSE(u,v) = 9.79\%$. Similar phenomenon could be found in other cases for other partition ratios and Reynolds numbers, which indicate that the overfitting of the proposed $\mathscr{M}_{FCN}$ model is much milder than that of the $\mathscr{M}_{CNN}$ model. Moreover, from tables ~\ref{tab:results_nmse_re_1000},~\ref{tab:results_nmse_re_500} and ~\ref{tab:results_nmse_re_100}, it is obvious that the metrics for our proposed $\mathscr{M}_{FCN}$ model are much lower than those for the $\mathscr{M}_{CNN}$ and $\mathscr{M}_{DNN}$ model, which indicate that our proposed $\mathscr{M}_{FCN}$ model have a better performance than the other two classical models with a relative small training dataset. This is because, the proposed $\mathscr{M}_{FCN}$ model could learn the features and the gradient information from neighbour nodes.}

By comparing the metrics of $\mathscr{M}_{CNN}$, $\mathscr{M}_{DNN}$ and $\mathscr{M}_{FCN}$ trained by 10\%, 30\% and 50\% training dataset, we find that with the reduction of training datasets, the difference between $\mathscr{M}_{CNN}$ and $\mathscr{M}_{DNN}$ training dataset $ E_{nmse} $ and testing dataset $ E_{nmse} $ is becoming larger, which means that the performances of the $\mathscr{M}_{CNN}$ and $\mathscr{M}_{DNN}$ are related to the size of training datasets, while the performace of our proposed FCN model is less affected by the size of training dataset.

\begin{figure*}
  \centerline{
    \includegraphics[width= 0.7\textwidth,trim=50 20 120 50,clip]{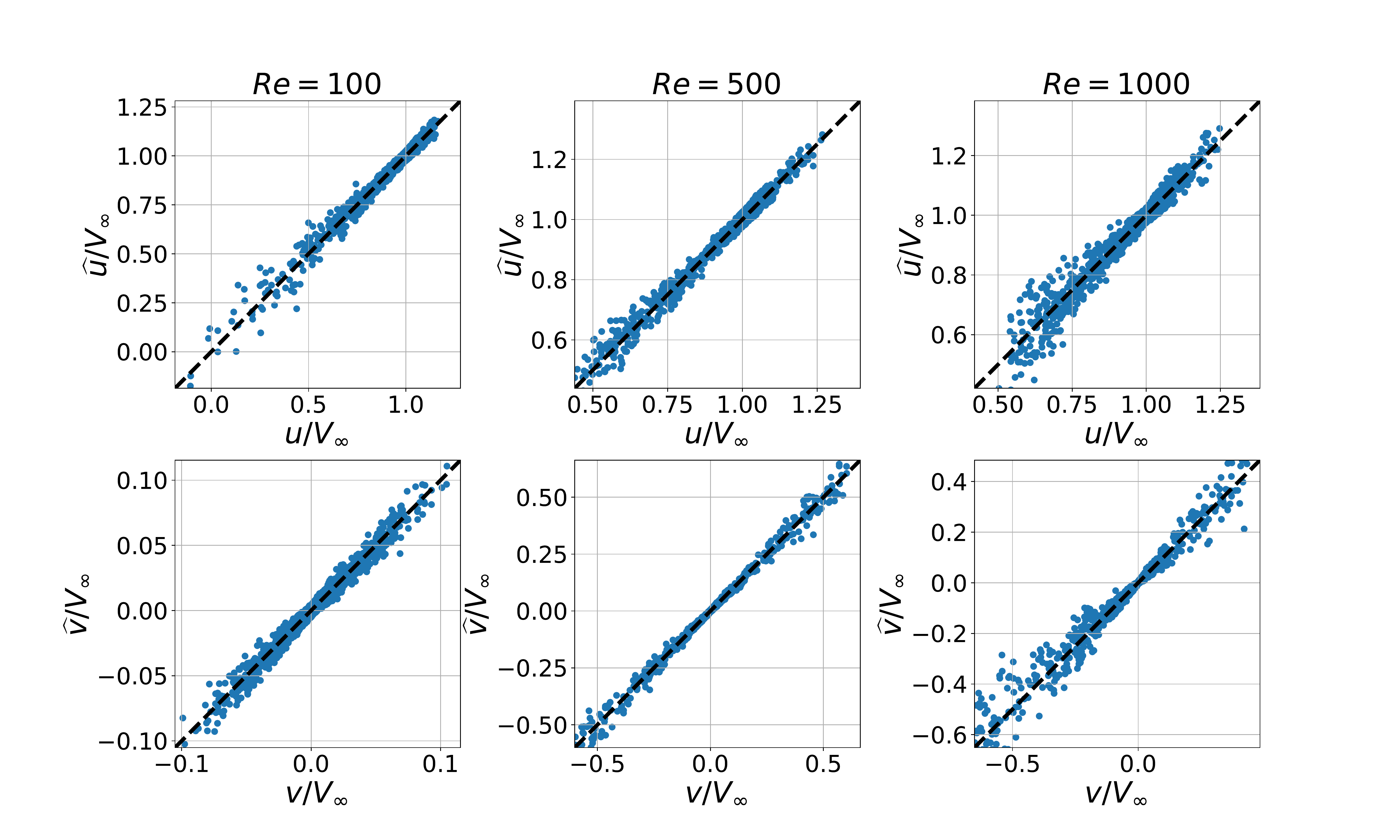}}
  \caption{\textcolor{black}{The comparison between the $u/{V_{\infty}}$, $v/{V_{\infty}}$ predicted by the proposed $\mathscr{M}_{FCN}$ models and those from the CFD results. The first column is for $Re = 100$, the second column is for $Re = 500$, and the third column is for $Re = 1000$.}} 
  \label{fig:result.uncertainty.reliability}
\end{figure*}

\textcolor{black}{Table \ref{tab:results_uncertainy_re_1000} shows the uncertainty assessment for the proposed $\mathscr{M}_{FCN}$ models trained by different datasets. The correlation coefficients, defined in Eq.~(\ref{eq:Metrics-CC}), in table \ref{tab:results_uncertainy_re_1000} for different cases are all close to $1$, which indicates that the predicted value has a strong correlation to the ground truth (the CFD results here). The $NMSE$ for all the cases are smaller than $5.86\%$ and the Bias for $u/{V_{\infty}}$ and $v/{V_{\infty}}$ are within $\pm0.0338$, which again shows the high accuracy our proposed FCN model.}
\textcolor{black}{Figure~\ref{fig:result.uncertainty.reliability} shows our proposed $\mathscr{M}_{FCN}$ model is fair in prdicting the flow field $u/{V_{\infty}}$, $v/{V_{\infty}}$ for different Reynolds number cases.}

\begin{table}
  \centering
  \begin{tabular}{cccccc}
  \\\hline
  Model (Re)                        & $PR$ & $ CC $   & $ NMSE(\%) $   & $Bias(u/V_{\infty})$ & $Bias(v/V_{\infty})$  \\\hline
  $ \mathscr{M}_{FCN} (Re = 100)  $ &10\%  & 0.9678   & 4.64\%         & +0.0020              & -0.0018               \\\
  $ \mathscr{M}_{FCN} (Re = 100)  $ &30\%  & 0.9867   & 3.59\%         & -0.0012              & +0.0002               \\\
  $ \mathscr{M}_{FCN} (Re = 100)  $ &50\%  & 0.9693   & 3.82\%         & -0.0034              & +0.0067               \\\hline
  $ \mathscr{M}_{FCN} (Re = 500)  $ &10\%  & 0.9576   & 3.84\%         & -0.0268              & -0.0422               \\\
  $ \mathscr{M}_{FCN} (Re = 500)  $ &30\%  & 0.9718   & 4.15\%         & -0.0021              & -0.0104               \\\
  $ \mathscr{M}_{FCN} (Re = 500)  $ &50\%  & 0.9641   & 4.39\%         & -0.0059              & +0.0014               \\\hline
  $ \mathscr{M}_{FCN} (Re = 1000) $ &10\%  & 0.9478   & 5.86\%         & -0.0338              & +0.0038               \\\
  $ \mathscr{M}_{FCN} (Re = 1000) $ &30\%  & 0.9597   & 4.72\%         & +0.0361              & -0.0229               \\\
  $ \mathscr{M}_{FCN} (Re = 1000) $ &50\%  & 0.9510   & 3.21\%         & +0.0272              & -0.0146               \\\hline
  \end{tabular}
  \caption{\textcolor{black}{The uncertainty assessment, includes the correlation coefficient (CC), the normalized mean square error ($NMSE$), and the Bias, for the proposed $\mathscr{M}_{FCN}$ models trained by different datasets. Here, the partition ratio (PR) means the proportion of the training dataset.} }
  {\label{tab:results_uncertainy_re_1000}}
\end{table}

\begin{figure*}
  \centerline{
    \includegraphics[width=0.75\textwidth]{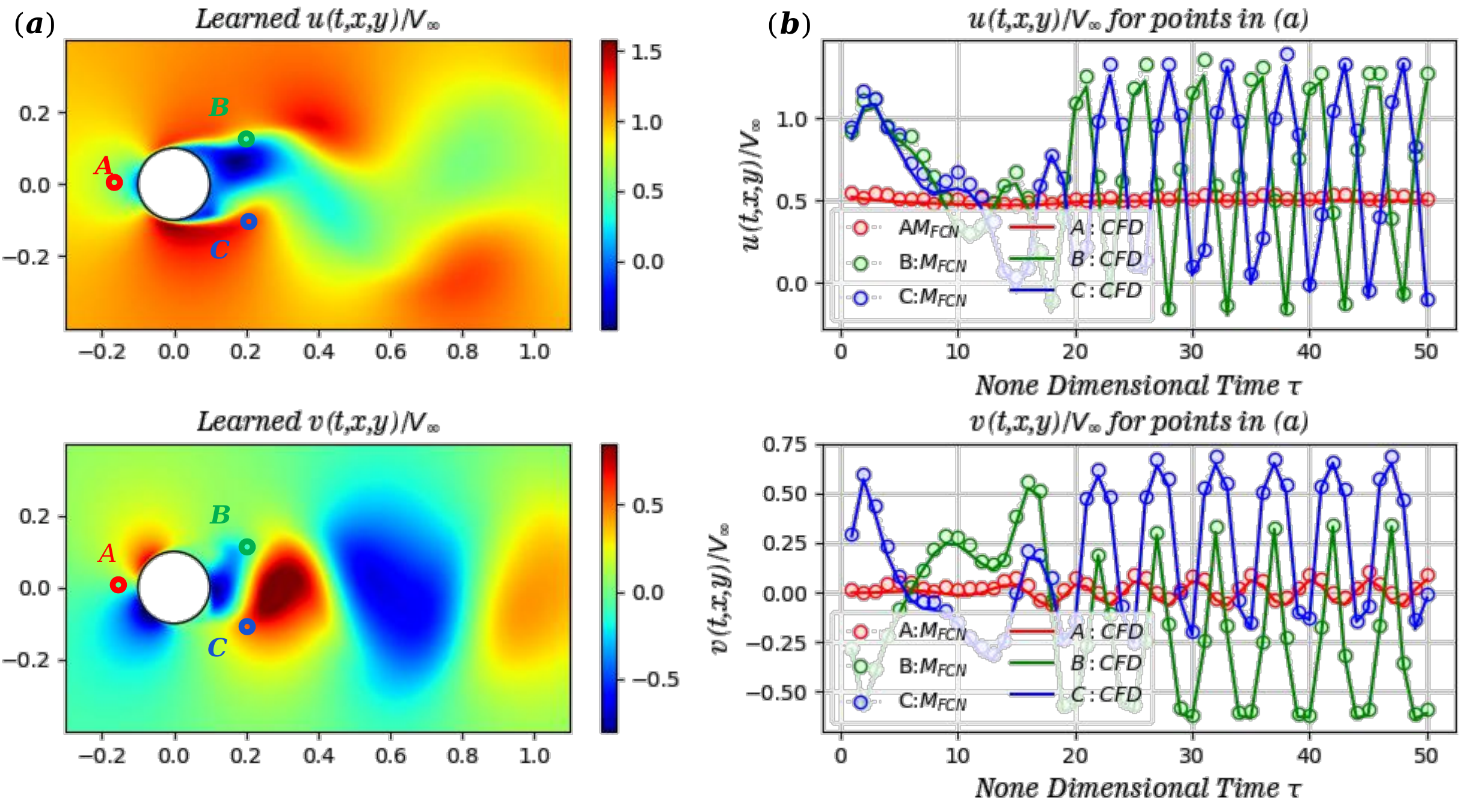}
    }
  \caption{The flow field completion results for (a) the velocity contours ($u/V_\infty,v/V_\infty$) at a typical instant and (b) the time variation of velocity  ($u/V_\infty,v/V_\infty$) at specific points $ A, B, C $ obtained by the FCN model.} 
  \label{fig:result.infinity.map}
\end{figure*}

In order to demonstrate the fitness/generality of the proposed FCN model, three observation points (A, B, and C) are selected from the cylinder flow field at $Re = 1000$ (see Fig.~\ref{fig:result.infinity.map} (a)) to check the variation of predicted flow features (velocity) against time. The comparison of the velocity variation against time obtained from the FCN model and extracted directly from the CFD is shown in Fig.~\ref{fig:result.infinity.map} (b). It is shown that the velocity variation predicted by the proposed FCN model compares well with the CFD data, and it has a strong generalization ability. Moreover, the error of the $\mathit{\mathscr{M}_{FCN}}$ does not vary with both the location of the observation point and the time.

\section{\label{sec:conclusion}Conclusion}

In this work, we introduced a novel model $\mathit{\mathscr{M}_{FCN}}$ based on GraphSAGE for
the flow field completion through using unstructured scattered data. 
The $\mathit{\mathscr{M}_{FCN}}$ was well designed to contain two GC layers and three SGA layers. The GC layers were introduced to take advantage of the properties of graph convolution neural networks, such as the internal physical law of the flow field (N-S equations). And the SGA layers were introduced to include the spatial gradient information while dealing with unstructured data. As we know, the experimental measurements of the flow field properties are usually conducted on sparsely scattered points, leading to unstructured data that are difficult to process with traditional machine learning algorithms (e.g. CNN-based models). 

To test the proposed FCN model, CFD simulation of a two-dimensional circular cylinder flow at different Reynolds numbers (Re = 100, Re = 500, Re = 1000) on the unstructured mesh was conducted here to provide training data set. The CFD results also served as the 'ground truth'. The relative error of the lift and drag coefficients, the NMSE of the two velocity components, as well as the CC and Bias of the velocity components were introduced to evaluate the performance of the proposed FCN model. 10\%, 30\%, and 50\% uniform randomly distributed scatter subtracted from the total dataset with and without unobserved regions have been used as training datasets. The comparison of experimental results from our proposed model together with two other traditional CNN and DNN-based models with CFD 'groud truth' showed the superiority of our FCN model in predicting the flow field feature and body force from incomplete flow measurements on unstructured mesh or scattered points. \textcolor{black}{The efficiency and accuracy of the proposed FCN model were less affected by decreasing the training dataset, and even 10\% of the whole dataset gave a reasonable prediction with a 5.86\% NMSE in the testing dataset for the case of $Re = 1000$. The NMSE for our proposed FCN model is much lower than those for the traditional CNN and DNN-based models. The output and input parameters of the FCN model show strong correlations and the Biases for the predicted flow velocity are minor.}  In a nutshell, this well-designed network and variable loss functions made the model being trained quickly and robustly. 

In summary, a novel neural network FCN has been proposed in this work to infer the fluid dynamics, including the flow field and the force acting on the body, from the incomplete data based on the graph convolution attention network. The FCN was designed to be capable of dealing with both structured data and unstructured data. The experimental results showed that our FCN model effectively utilizes the existing flow field information and the gradient information simultaneously, giving a better prediction of the flow field and body force than the traditional CNN-based and DNN-based models.

\begin{acknowledgments}
  \emph{This work has received funding from the European Union's Horizon 2020 research and innovation programme under the Marie Sklodowska-Curie grant agreement No.765579. This work is funded by the Leverhulme Trust, Grant Ref ECF-2018-727. Their support is gratefully acknowledged.}

\end{acknowledgments}

\section*{Data Availability Statement}

The data that support the findings of this study are available from the corresponding author upon reasonable request.

\nocite{*}
\bibliography{aippaper}
\end{document}